\documentclass[aps,prb,twocolumn,groupedaddress,superscriptaddress,showpacs]{revtex4-1}
\usepackage{graphicx}
\usepackage{amsmath}
\usepackage{hyperref}
\usepackage{textcomp}
\hypersetup{colorlinks=true, citecolor=blue, urlcolor=blue, linkcolor=blue}

\newcommand {\SPS}    {Sn$_2$P$_2$S$_6$ }
\newcommand {\PPS}    {Pb$_2$P$_2$S$_6$ }
\newcommand {\PPSe}    {Pb$_2$P$_2$Se$_6$ }

\newcommand {\PSPS}    {(Pb$_y$Sn$_{1-y}$)$_2$P$_2$S$_6$ }
\newcommand {\PSPSe}    {(Pb$_y$Sn$_{1-y}$)$_2$P$_2$Se$_6$ }
\newcommand {\SPSS}    {Sn$_2$P$_2$(Se$_x$S$_{1-x}$)$_6$ }

\begin{document}

\title{Quantum paraelectric state and critical behavior in Sn(Pb)$_2$P$_2$S(Se)$_6$ ferroelectrics}

\author{I.~Zamaraite}
\affiliation{Faculty of Physics, Vilnius University, Sauletekio 9, 10222 Vilnius, Lithuania}
\author{V.~Liubachko}
\affiliation{Institute for Solid State Physics and Chemistry, Uzhhorod University, Pidgirna Str. 46, Uzhhorod, 88000, Ukraine}
\affiliation{Departamento de Física Aplicada I, Escuela de Ingeniería de Bilbao, Universidad del País Vasco UPV/EHU, Plaza Torres Quevedo 1, 48013 Bilbao, Spain}
\author{R.~Yevych}
\affiliation{Institute for Solid State Physics and Chemistry, Uzhhorod University, Pidgirna Str. 46, Uzhhorod, 88000, Ukraine}
\author{A.~Oleaga}
\affiliation{Departamento de Física Aplicada I, Escuela de Ingeniería de Bilbao, Universidad del País Vasco UPV/EHU, Plaza Torres Quevedo 1, 48013 Bilbao, Spain}
\author{A.~Salazar}
\affiliation{Departamento de Física Aplicada I, Escuela de Ingeniería de Bilbao, Universidad del País Vasco UPV/EHU, Plaza Torres Quevedo 1, 48013 Bilbao, Spain}
\author{A.~Dziaugys}
\affiliation{Faculty of Physics, Vilnius University, Sauletekio 9, 10222 Vilnius, Lithuania}
\author{J.~Banys}
\affiliation{Faculty of Physics, Vilnius University, Sauletekio 9, 10222 Vilnius, Lithuania}
\author{Yu.~Vysochanskii}
\affiliation{Institute for Solid State Physics and Chemistry, Uzhhorod University, Pidgirna Str. 46, Uzhhorod, 88000, Ukraine}

\date{\today}

\begin{abstract}
The dipole ordering in Sn(Pb)$_2$P$_2$S(Se)$_6$ materials may be tuned by chemical substitution realizing a ferroelectric quantum phase transition and quantum glassy or relaxor type phenomena on different parts of the phase diagram. The introduction of Ge impurity increases the temperature of the phase transitions and initiates a more pronounced Ising type critical anomaly in  \SPS crystal, does not shift the coordinate of the Lifshitz point $x_{\textrm {LP}}$ in \SPSS mixed crystals, induces the appearance of a ferroelectric phase transition in quantum paraelectrics \PPS and inhomogeneous polar ordering in  (Pb$_{0.7}$Sn$_{0.3}$)$_2$P$_2$S(Se)$_6$ crystals. For \PPS crystal, the real part of the dielectric susceptibility in the quantum critical regime varies as $1/T^2$ instead of the expected $1/T^3$ behavior for uniaxial materials. This can be partially explained by a screening phenomenon in the semiconductor materials of the Sn(Pb)$_2$P$_2$S(Se)$_6$ system, which weakens the long range electric dipole interactions, and also provides, at high temperatures, a critical behavior near the Lifshitz point (studied by thermal diffusivity) similar to the one predicted in the case of systems with short range interactions.
At low temperatures, a quantum critical behavior in \PPS crystal can be established by the nonlinear coupling between polar and antipolar fluctuations. An increase in thermal conductivity is induced by Ge impurity in \PPS crystal, which is explained through the weakening of the acoustic phonons resonance scattering by soft optic phonons because of the appearance of ferroelectric phase polar clusters.
\end{abstract}

\maketitle

\section{Introduction}
It was earlier found\cite{b2019_1} that Ge doping shifts the second order phase transition (P2$_1$/c$\leftrightarrow$Pc) in uniaxial ferroelectric \SPS crystal toward higher temperatures. The increase of the phase transition temperature under the influence of Ge impurities is also known for Pb$_{1-x}$Ge$_x$Te and Sn$_{1-x}$Ge$_x$Te ferroelectrics,\cite{b2019_2} what demonstrates the universal property of Ge impurities in tin or lead containing hosts of elevating the ferroelectric phase transition temperature. By means of X-ray photoelectron spectroscopy, together with first-principles calculations of electronic spectra, it was found\cite{b2019_3} that the germanium impurity in \SPS improves the stereoactivity of the cation sublattice. In \SPS ferroelectrics the Sn$^{2+}$ cations stereoactivity and the $P^{4+} + P^{4+} \leftrightarrow P^{3+} + P^{5+}$ charge disproportionation is related to the nature of the second order phase transition with mixed displacive-order/disorder character.\cite{b2019_4,b2019_5}

\begin{figure}[!htb]
\includegraphics*[width=\columnwidth]{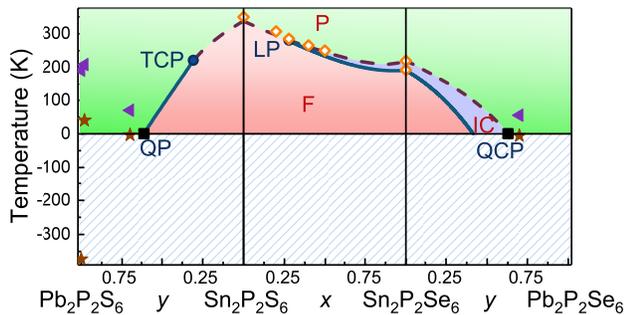}
 \caption{Phase diagram temperature-composition for mixed crystals in Sn(Pb)$_2$P$_2$S(Se)$_6$ system.\cite{b2019_6} Phase transitions in crystals with Ge impurity denote orange rhombs.\cite{b2019_1,b2019_16,b2019_36}  For crystals with quantum paraelectric state determined by Eq.~(\ref{eq1}) values of $T_{\textrm C}$ (brown stars) and $T_1$ (violet triangles) are shown. Filled circle denote tricritical point (TCP). Black squares denote quantum point (QP) for the line of first order paraelectric-ferroelectric transitions, and quantum critical point (QCP) for the line of second order paraelectric-incommensurate transitions. Paraelectric (P), ferroelectric (F) and incommensurate (IC) phases are also shown. First order phase transion lines shown by solid lines, second order ones --- by dashed lines. \label{fig1}}
\end{figure}

While germanium can only be introduced up to a certain quantity,\cite{b2019_1} lead can completely substitute tin in \PSPS and \PSPSe continuous solid solutions.\cite{b2019_6} The replacement of tin by lead in the cationic sublattice induces the lowering of the phase transition temperature (see Fig.~\ref{fig1}), changing its character to a discontinuous transition across the tricritical point (TCP), and stabilizing the paraelectric phase in the ground state (at $T = 0$~K) for $y > 0.61$.\cite{b2019_7} The addition of Pb has the effect of diluting the stereoactivity as it weakens the bonding hybridization responsible for ferroelectricity. Formally, the introduction of lead atoms creates a “chemical pressure” with similar effects to the mechanical pressure.\cite{b2019_5} The properties of the \PSPS mixed crystals have been described\cite{b2019_7} within the framework of Blume-Emery-Griffith (BEG) model\cite{b2019_8,b2019_9} taking into account the presence of random fields created when substituting tin by lead.

The substitution of S by Se in the anionic sublattice provokes the appearance of an incommensurate (IC) phase at the Lifshitz point (LP) for $x > x_{\textrm {LP}} \approx 0.28$ in \SPSS solid solutions.\cite{b2019_10} The line of tricritical points meet with the line of Lifshitz points at the tricritical Lifshitz point on the $T - x - y$ phase diagram with an interesting topology.\cite{b2019_11}

In mixed crystals \PSPS and \PSPSe at $y > 0.61$ and $y > 0.65$, respectively, the paraelectric phase is stable in the ground state,\cite{b2019_6,b2019_12} and pure compounds \PPS and \PPSe are quantum paraelectrics. The quantum paraelectric state is manifested in the \PPS crystal by the growth of the dielectric susceptibility while cooling down to 0~K.\cite{b2019_5} A similar state also appears in the \SPS crystal at hydrostatic pressure $p > 1.5$~GPa.\cite{b2019_13,b2019_14}

The Pb$^{2+}$ cations have a smaller stereoactivity compared with Sn$^{2+}$ and this determines the suppression of ferroelectricity while substituting tin by lead.\cite{b2019_5,b2019_15} In the case of Ge$^{2+}$ cations, the stereoactivity is bigger, which provokes a temperature rise in the ferroelectric phase transition in \SPS crystals with germanium impurity.\cite{b2019_1,b2019_16} For \SPSS solutions, both S by Se and Sn by Ge replacements increase the crystal lattice covalence, and the critical anomaly near the LP becomes sharper.\cite{b2019_36}

In the case of the binary compounds GeS, SnS, PbS or GeS, GeSe, GeTe, it was demonstrated that it is mostly the energy difference between $s$ orbitals of metal atoms and $p$ orbital of chalcogen atoms which determines the stereoactivity of the cations and the crystal lattice covalency.\cite{b2019_17} For ternary compounds, for example BiNiO$_3$, it was demonstrated\cite{b2019_18} that the energy positions of the valence orbitals of the two cations Bi$^{4+}$ and Ni$^{2+}$ are also important. Therefore, for compounds of the Sn(Pb)$_2$P$_2$S(Se)$_6$ system with two types of cations (tin or lead metals and phosphorous), a more complex role of the electron valence orbitals hybridization can also be important.\cite{b2019_5,b2019_15,b2019_19} It was found \cite{b2019_5} that Sn by Pb substitution changes the local potential for spontaneous polarization fluctuation at almost constant intercell interactions. On S by Se substitution, on the contrary, the intersite interaction is changed.
 
In previous investigations \cite{b2019_5,b2019_6,b2019_12} the influence of Sn$\rightarrow$Pb and S$\rightarrow$Se substitutions on the phase transition from the paraelectric phase into the ferroelectric one has been analyzed. Here we will pay attention to understanding the germanium impurity influence on the phase transitions and the quantum paraelectric state in different segments of the Sn(Pb)$_2$P$_2$S(Se)$_6$ ferroelectrics phase diagram. The temperature dependence of dielectric susceptibility and thermal diffusivity are analyzed with the use of the quantum anharmonic oscillators (QAO) model \cite{b2019_5,b2019_20} for the calculation of the spontaneous polarization fluctuations spectra in the local three-well potential. The appearance of the quantum critical behavior in \PPS and (Pb$_{0.98}$Ge$_{0.02}$)$_2$P$_2$S$_6$ crystals, in mixed \PSPS and \PSPSe crystals with 5\% of Ge impurity is investigated, together with the appearance of polar ordering at low temperatures induced by germanium.

\section{Experimental data}

We have investigated the temperature dependence of the dielectric susceptibility with a HP4284 precision LCR meter at temperatures from 300 to 20~K during the cooling cycle at a rate of about 1~K/min, and at frequencies ranging from 20 to 1~MHz.\cite{b2019_21,b2019_22} The thermal diffusivity $D$ measurements have been performed by a high resolution $ac$ photopyroelectric calorimetry technique in the standard back detection configuration. A closed cycle He cryostat operated in cooling and heating modes has been used.\cite{b2019_16} Ge-doped single crystals were obtained by vapor-transport method in a quartz tube using SnI$_2$ as a transport agent. The synthesis of the starting material in the polycrystalline form was carried out using high-purity (99.999\%) elements.\cite{b2019_16,b2019_23} The samples were characterized and oriented by X-ray diffraction technique. For complex dielectric susceptibility measurements, the monocrystal plates with the thickness of about 2~mm and plane parallel faces around 15~mm$^2$ with silver paste electrodes on polar (100) faces were prepared. For thermal diffusivity measurements all samples have been prepared in the form of thin plane-parallel slabs with thicknesses in a range of 0.500-0.550 mm and whose faces were cut in the monoclinic symmetry plane.

\begin{figure}[!htb]
\includegraphics*[width=0.95\columnwidth]{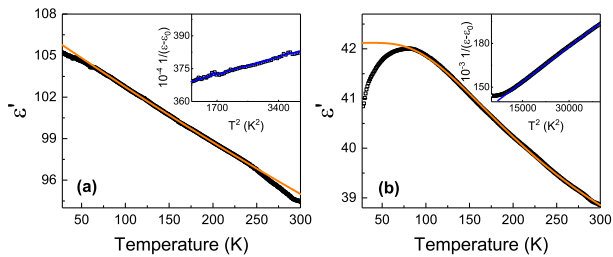}
 \caption{Temperature dependence of (a) the real part of dielectric susceptibility for \PPS crystal at 100~kHz and (b) (Pb$_{0.98}$Ge$_{0.02}$)$_2$P$_2$S$_6$ at 10~kHz. Orange lines are the fitting of Eq.~(\ref{eq1}). The inset shows the $\varepsilon(T)^{-1} \sim T^2$ behavior (blue lines). \label{fig2}}
\end{figure}

\begin{figure}[!htb]
\begin{minipage}[c]{0.3\columnwidth}
\includegraphics*[width= 1.9in]{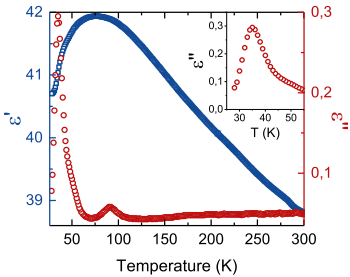}
\end{minipage}\hfill
  \begin{minipage}[c]{0.41\columnwidth}
   \caption{Temperature dependence of dielectric susceptibility real $\varepsilon '$ (blue squares) and imaginary $\varepsilon ''$(red circles) parts at 100~kHz for (Pb$_{0.98}$Ge$_{0.02}$)$_2$P$_2$S$_6$ crystal.\label{fig3}}
  \end{minipage}
\end{figure}

\begin{figure}[!htb]
\includegraphics*[width=0.9\columnwidth]{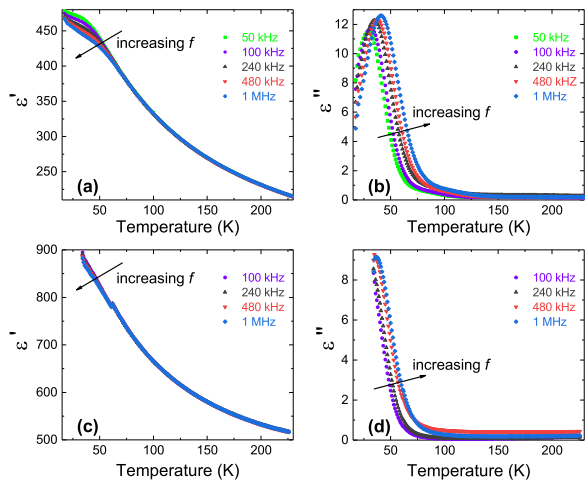}
 \caption{Temperature dependence of the real and imaginary parts of dielectric susceptibility at different frequencies $f$ of applied field for (a, b) (Pb$_{0.7}$Sn$_{0.25}$Ge$_{0.05}$)$_2$P$_2$S$_6$ and (c, d) (Pb$_{0.7}$Sn$_{0.25}$Ge$_{0.05}$)$_2$P$_2$Se$_6$ crystals. \label{fig4}}
\end{figure}

\begin{figure}[!htb]
\includegraphics*[width=0.9\columnwidth]{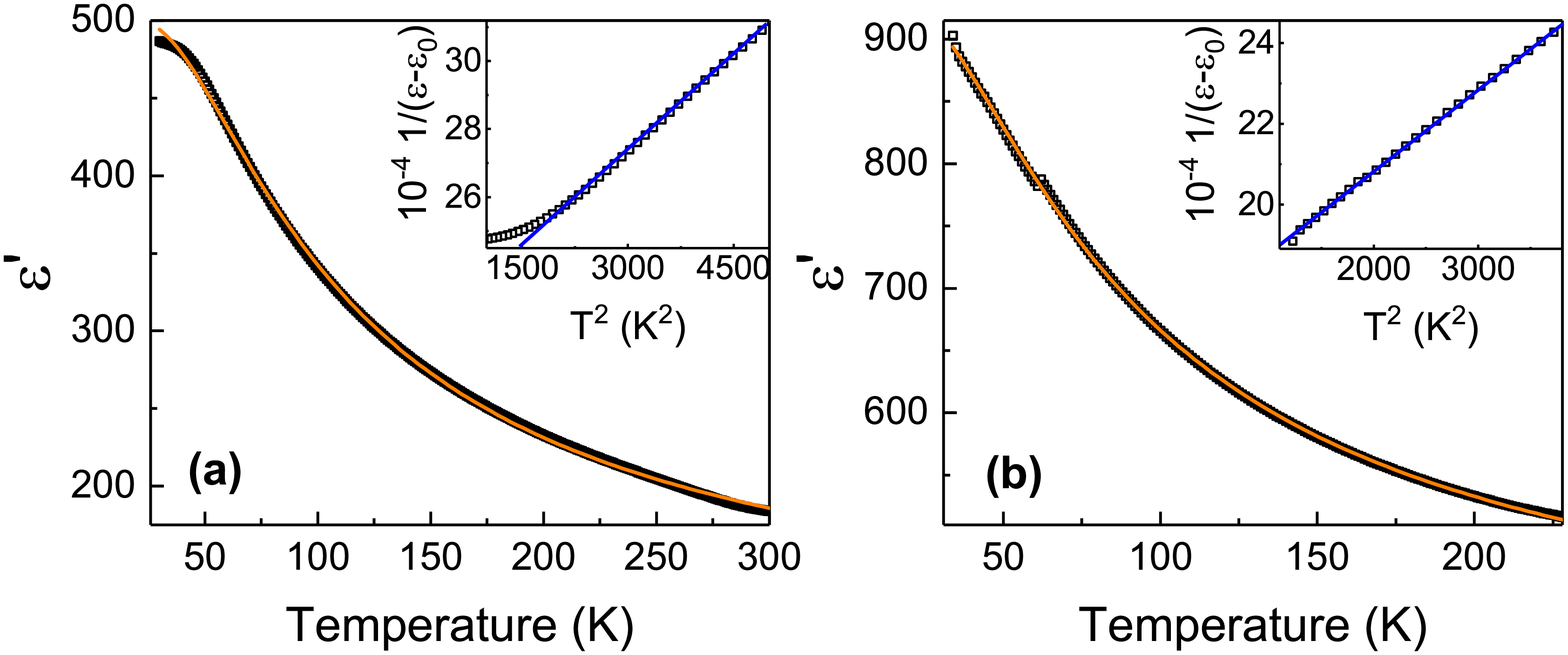}
 \caption{Temperature dependence of dielectric susceptibility real part of (a) (Pb$_{0.7}$Sn$_{0.25}$Ge$_{0.05}$)$_2$P$_2$S$_6$  and (b) (Pb$_{0.7}$Sn$_{0.25}$Ge$_{0.05}$)$_2$P$_2$Se$_6$ at 10~kHz; orange solid curve is the fitting according to the Barret`s equation~(\ref{eq1}). The inset shows the reciprocal dielectric susceptibility as a function of the squared temperature (blue lines). \label{fig5}}
\end{figure}

For \PPS crystal the $\varepsilon '(T)$ dependence shows monotonic rise at cooling till 20~K (see Fig.~\ref{fig2}) with some flattening below 50~K. At lead by germanium substitution (about 2\%), a clear maximum of $\varepsilon ''(T)$ near 35~K (at 100~kHz frequency) is observed (see Fig.~\ref{fig3})  and below this temperature $\varepsilon '(T)$ displays frequency dependency in the range between 1~kHz and 1~MHz. In the case of (Pb$_{0.7}$Sn$_{0.25}$Ge$_{0.05}$)$_2$P$_2$S$_6$ and (Pb$_{0.7}$Sn$_{0.25}$Ge$_{0.05}$)$_2$P$_2$Se$_6$ mixed crystals, $\varepsilon '(T)$ and $\varepsilon ''(T)$ anomalies also appear in the temperature region of 20--50~K (see Fig.~\ref{fig4}, \ref{fig5}). These anomalies are induced by germanium impurity.

By the previous dielectric investigations \cite{b2019_12} it was shown that for \PSPS mixed crystals with $y = 0.61$ and $y = 0.66$ in the paraelectric phase $\varepsilon '(T) \sim (T - T_{\textrm C})^{-2}$.  Such dependence was attributed to glassy-like dielectric susceptibility behavior in \PSPS mixed crystals with coexisting paraelectric and ferroelectric states. We have shown that such temperature dependence of the dielectric susceptibility can be attributed to the appearance of the spontaneous polarization quantum fluctuations at low temperatures in the crystals of the investigated system.

For \PPS crystal the real part of the dielectric susceptibility increases monotonously with decreasing temperature and in the measured temperature range the saturation behavior is not observed [see Fig.~\ref{fig2}(a)]. In the quantum critical regime the usual Curie-Weiss law of the inverse dielectric susceptibility $1/\varepsilon(T) \sim T$ changes into $1/\varepsilon(T) \sim T^2$.\cite{b2019_25,b2019_26} That is the most prominent criterion for quantum critical behavior. For \PPS the inverse dielectric susceptibility $1/\varepsilon(T)$ exhibits the expected non-classical $T^2$ temperature dependence over the temperature range 50 K--250~K. 

In order to describe the temperature dependencies of the dielectric susceptibility of quantum paraelectrics, Barrett‘s equation can be used:
 \begin{equation}
\varepsilon(T)=\frac{C}{(\frac{T_1}{2})\coth(\frac{T_1}{2T})-T_{\textrm C}}+\varepsilon_0, \label{eq1}
\end{equation}
\noindent where $C$ is the Curie-Weiss constant, $T_{\textrm C}$ is the classical paraelectric Curie temperature, $\varepsilon_0$ is a temperature independent constant  and $T_1$ is  the dividing point between the low temperatures where quantum effects are important so $\varepsilon(T)$ deviates from Curie-Weiss law, and the high temperature region where a classical approximation and Curie-Weiss law are valid.\cite{b2019_26,b2019_42}

In many cases, $T_{\textrm C} \le 0$~K, and the material does not undergo a ferroelectric phase transition at any finite temperature. When $T_{\textrm C}$ is finite and $T_{\textrm C} < T_1$, the quantum fluctuations break the long range ferroelectric order and stabilize the quantum paraelectric state in the sample. Probable ferroelectric transition occurs at $T_{\textrm C}$.\cite{b2019_27} According to dielectric data of \PPS [see Fig.~\ref{fig2}(a)] deviation from Barrett`s equation starts around 75~K. The obtained parameter values ($T_1 \approx 190$~K and $T_{\textrm C} \approx - 370$~K) for \PPS crystal demonstrate that the material does not undergo a ferroelectric phase transition at any finite temperature.

As was mentioned above, when Pb substitutes Sn in \SPS type crystals, the hybridization of anion and cation sublattices electron orbitals becomes weaker, reducing the phase transition temperature. On the other hand, Ge dopant plays an opposite role: it enhances the total stereoactivity of the cation sublattice in the crystal. Small amount of impurities in quantum paraelectrics could induce ferroelectricity.\cite{b2019_28,b2019_29} So, it is possible that germanium impurities can affect quantum paraelectric state of Pb$_2$P$_2$S$_6$. Figure~\ref{fig2}(b) shows the temperature dependence of the real part of dielectric susceptibility for (Pb$_{0.98}$Ge$_{0.02}$)$_2$P$_2$S$_6$ crystal and confirms a non-classical $T^2$ behavior of the inverse dielectric susceptibility. For this, the temperature dependence of the real part of dielectric susceptibility for crystal doped by germanium is fitted by the Barret’s equation (\ref{eq1}) giving temperatures $T_1\approx 200$~K and $T_{\textrm C}\approx 40$~K. Since $T_{\textrm C} < T_1$ for (Pb$_{0.98}$Ge$_{0.02}$)$_2$P$_2$S$_6$, it could be concluded that the long-range ferroelectric order in the sample doped by Ge is broken due to quantum fluctuations below 200~K, and a probable ferroelectric transition occurs in the temperature region between 40~K and 80~K (see Fig.~\ref{fig3}). Doping with germanium decreases the real part of susceptibility below 80~K deviating from Barrett`s fit [see Fig.~\ref{fig2}(b)]. 

The peak of the real part of the dielectric susceptibility is broad. Moreover, there are two peaks of the imaginary part of the dielectric susceptibility with a frequency dispersive behavior, and the temperatures of the loss peaks are around 50~K and 100~K at 100~kHz (see Fig.~\ref{fig3}). Obviously, this is related to compositional fluctuations in (Pb$_{0.98}$Ge$_{0.02}$)$_2$P$_2$S$_6$ crystal. Also, a fast-enough dynamics of local dipoles, and slower dynamics of noninteracting ones, or weakly interacting nanoclusters, can determine the broadness of the phase transition induced by Ge impurity with related frequency--temperature anomalous behavior of dielectric susceptibility that is similar to the one observed in the case of a crossover between dipole glass and ferroelectric relaxor.\cite{b2019_28,b2019_29}

As was already mentioned, for \PSPS mixed crystals with compositions  $y =  0.61$ and $y =  0.66$, which are close to the transition at zero temperature from a polar phase to a paraelectric one, the dielectric susceptibility demonstrates the quantum critical behavior with $T_{\textrm C}\approx  35$~K and 20~K, respectively.\cite{b2019_12} We have investigated the influence of Ge dopants on the quantum paraelectric state of \PPS type compounds by studying of (Pb$_{0.7}$Sn$_{0.25}$Ge$_{0.05}$)$_2$P$_2$S$_6$ and (Pb$_{0.7}$Sn$_{0.25}$Ge$_{0.05}$)$_2$P$_2$Se$_6$ samples. In these crystals the Sn$^{2+}$ sites of pure \SPS were codoped with two different impurities (Pb$^{2+}$ and Ge$^{2+}$) which have very different influences on the phase transitions.  It is important to realize that Sn substitution has the strongest effect because the ferroelectric phase transition is induced by the stereoactivity of the Sn$^{2+}$ cation $5s^2$ electron lone pair. 

The temperature dependence of the real part of the dielectric susceptibility for (Pb$_{0.7}$Sn$_{0.25}$Ge$_{0.05}$)$_2$P$_2$S$_6$ and (Pb$_{0.7}$Sn$_{0.25}$Ge$_{0.05}$)$_2$P$_2$Se$_6$ crystals is shown in Fig.~\ref{fig4}. Susceptibility $\varepsilon '$ increases continuously with decreasing temperature from room temperature till 20~K. The dielectric losses have maximum at low temperature, around 40~K at frequency 1~MHz. The inverse dielectric permittivity $1/\varepsilon(T)$ exhibits the expected non-classical $T^2$ temperature dependence not only in the case of doped (Pb$_{0.98}$Ge$_{0.02}$)$_2$P$_2$S$_6$ sample (see Fig.~\ref{fig2}), but it is also observed in mixed crystals (Pb$_{0.7}$Sn$_{0.25}$Ge$_{0.05}$)$_2$P$_2$S$_6$ and (Pb$_{0.7}$Sn$_{0.25}$Ge$_{0.05}$)$_2$P$_2$Se$_6$ (see Fig.~\ref{fig5}). From this follows that the ferroelectric quantum critical behavior is relatively insensitive to quenched disorder in doped samples and mixed crystals. 

By fitting the experimental data of Fig.~\ref{fig5}(a) to Barrett`s equation~(\ref{eq1}) it was determined the next parameters:  $T_1 \approx 70$~K, $T_{\textrm C}\approx - 4$~K, and $C \approx 30670$~K. The observed temperature behavior of the dielectric susceptibility demonstrates that (Pb$_{0.7}$Sn$_{0.25}$Ge$_{0.05}$)$_2$P$_2$S$_6$ crystal obviously undergoes some inhomogeneous polar ordering at very low temperatures.

Similarly, the temperature dependence of the dielectric susceptibility $\varepsilon '(T)$  for the (Pb$_{0.7}$Sn$_{0.25}$Ge$_{0.05}$)$_2$P$_2$Se$_6$ crystal is shown in Fig.~\ref{fig5}(b). On cooling from 300 till 20~K both $\varepsilon '$ and $\varepsilon ''$ increase, their frequency dispersion more clearly appears below 100~K.\cite{b2019_22} By fitting to the Barret`s equation~(\ref{eq1}) [see Fig.~\ref{fig5}(b)] it was found that $T_1 \approx 55$~K, $T_{\textrm C}\approx - 6$~K, and $C \approx 34260$~K. It is seen that in the selenide mixed crystal (Pb$_{0.7}$Sn$_{0.25}$Ge$_{0.05}$)$_2$P$_2$Se$_6$ the germanium impurity induces inhomogeneous polar ordering at lower temperatures similarly to the case of the sulfide analogue.

As a whole, according to the results of the dielectric investigations, it can be concluded that \PPS crystals exhibit a quantum paraelectric state. The introduction of small amounts of germanium dopant provokes the appearance of the ferroelectric phase. In mixed crystals a very inhomogeneous polar ordering (like dipole glassy or relaxor state) appears below approximately 100~K.

We can see that the $\varepsilon '(T)$ dependence of Eq.~(\ref{eq1}), considering quantum fluctuations for \PPS compound, predicts the value of $T_{\textrm C}$ in accordance with $T_0(y)$ dependence for \PSPS mixed crystals (see Fig.~\ref{fig1}). At this, the value of $T_1$ is strongly suppressed in mixed crystals --- from 190~K in \PPS and 207~K in the case of (Pb$_{0.98}$Ge$_{0.02}$)$_2$P$_2$S$_6$ to $T_1 \approx 70$~K in the solution with $y = 0.7$. Such decrease of the crossover temperature $T_1$ from classic to quantum fluctuations behavior can be interpreted as the manifestation of quantum coherence destruction for the electronic component of spontaneous polarization fluctuations, that are determined by phosphorous cations $P^{4+} + P^{4+} \leftrightarrow P^{3+} + P^{5+}$ charge disproportionation. The electronic contribution to spontaneous polarization is connected to the coherent state of polaronic excitons --- small hole polarons in SnPS$_3$ structural groups are coupled with small electronic polarons in nearest SnPS$_3$ structural groups.\cite{b2019_5,b2019_30} Obviously such polaronic excitons are strongly bounded by defects in mixed crystals what preserves the development of quantum fluctuations when lowering the temperature.

\begin{figure}[!htb]
\includegraphics*[width=0.9\columnwidth]{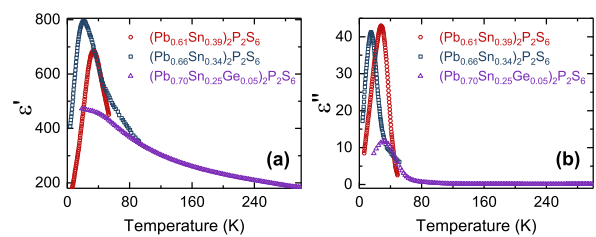}
 \caption{Temperature dependence of dielectric susceptibility (a) real and (b) imaginary parts for crystals (Pb$_{0.7}$Sn$_{0.25}$Ge$_{0.05}$)$_2$P$_2$S$_6$. Data [\onlinecite{b2019_12}] for (Pb$_{0.61}$Sn$_{0.39}$)$_2$P$_2$S$_6$ and (Pb$_{0.66}$Sn$_{0.34}$)$_2$P$_2$S$_6$ mixed crystals are also shown for comparison. \label{fig6}}
\end{figure}

In Fig.~\ref{fig6} the temperature dependencies of real and imaginary parts of dielectric susceptibility are compared for (Pb$_{0.7}$Sn$_{0.25}$Ge$_{0.05}$)$_2$P$_2$S$_6$ crystal in comparison with data [\onlinecite{b2019_12}] for (Pb$_{0.66}$Sn$_{0.34}$)$_2$P$_2$S$_6$ and (Pb$_{0.61}$Sn$_{0.39}$)$_2$P$_2$S$_6$ mixed crystals. We can see that the dielectric anomalies induced by the germanium impurity are smeared similarly to the observed anomalies in the case of \PSPS solid solutions with lead concentration near the threshold value of $y$. In all three samples a complex thermal evolution of the provoked inhomogeneous polarization occurs on cooling below 100~K.

The thermal properties of \SPS and (Sn$_{0.95}$Ge$_{0.05}$)$_2$P$_2$S$_6$, Sn$_2$P$_2$(Se$_{0.28}$S$_{0.72}$)$_6$ and (Sn$_{0.95}$Ge$_{0.05}$)$_2$P$_2$(Se$_{0.28}$S$_{0.72}$)$_6$, \PPS and (Pb$_{0.98}$Ge$_{0.02}$)$_2$P$_2$S$_6$, (Pb$_{0.7}$Sn$_{0.25}$Ge$_{0.05}$)$_2$P$_2$S$_6$ and (Pb$_{0.7}$Sn$_{0.25}$Ge$_{0.05}$)$_2$P$_2$Se$_6$ single crystals have been studied  by means of $ac$ photopyroelectric calorimetry, measuring the thermal diffusivity.\cite{b2019_16,b2019_31,b2019_34,b2019_35,b2019_36}  Thermal conductivity $k$ has been retrieved by combining the experimental thermal diffusivity $D$ and the calculated background heat capacity $c$ for Sn$_2$P$_2$S$_6$, \PPS and the experimental one for Sn$_2$P$_2$Se$_6$, Pb$_2$P$_2$Se$_6$.\cite{b2019_24,b2019_49} The detailed procedure to obtain the thermal conductivity is explained elsewhere.\cite{b2019_48}

\begin{figure}[!htb]
\includegraphics*[width=\columnwidth]{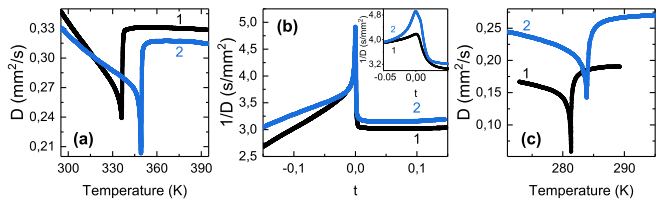}
 \caption{Temperature dependence of \SPS (1) and (Sn$_{0.95}$Ge$_{0.05}$)$_2$P$_2$S$_6$ (2) crystals (a) thermal diffusivity $D$ and (b) their reciprocal value as a function of reduced temperature $t = (T-T_{\textrm c})/T_{\textrm c}$; (c) thermal diffusivity $D$ temperature anomalies near the Lifshitz point in Sn$_2$P$_2$(Se$_{0.28}$S$_{0.72}$)$_6$ (1) and (Sn$_{0.95}$Ge$_{0.05}$)$_2$P$_2$(Se$_{0.28}$S$_{0.72}$)$_6$ (2) crystals.\cite{b2019_16,b2019_31,b2019_34,b2019_35} \label{fig7}}
\end{figure}

For \SPS crystal, the germanium dopant shifts the temperature of continuous ferroelectric transition upwards and sharpens the critical anomaly of thermal diffusivity (see Fig.~\ref{fig7}).\cite{b2019_16} The introduction of germanium impurity into \SPS crystal lattice increases the dip of the thermal diffusivity anomaly near the second order phase transition, which becomes a little broader than in the case of the nominally pure crystal [see Fig.~\ref{fig7}(b)] and therefore, it was not possible to perform fittings with great accuracy.\cite{b2019_16,b2019_34}

For the Lifshitz point composition Sn$_2$P$_2$(Se$_{0.28}$S$_{0.72}$)$_6$, the critical index $\alpha \approx 0.34$ and the critical amplitudes ratio $A^+/A^- \approx 0.42$ were observed.\cite{b2019_35} The introduction of Ge increases the critical temperature from 281.3 K to 284~K [see Fig.~\ref{fig7}(c)] but changes neither the character of the transition nor the universality class, as it is uniaxial Lifshitz class with critical exponent $\alpha \approx 0.25$ and $A^+/A^- \approx 0.49$.\cite{b2019_31} Such values agree with the theoretical ones estimated for a Lifshitz system without considering strong dipolar interactions. In the case of the Lifshitz point in uniaxial ferroelectrics only small multiplicative corrections to mean field behavior are expected\cite{b2019_37} in the critical region. Hence, both the critical exponent and the amplitude ratio values observed for the Sn$_2$P$_2$(Se$_{0.28}$S$_{0.72}$)$_6$ mixed crystal lead to the conclusion that long-range interactions do not have a strong influence on the critical behavior in this system. This can be related to the partial screening of the dipole-dipole interaction by charge carriers in the \SPSS ferroelectric semiconductors. 

In order to determine the thermal conductivity $k$ of the investigated samples, thermal diffusivity data $D$ have been combined with heat capacity data $c$ using the following equation
 \begin{equation}
k=cD.\label{eq2}
\end{equation}

\begin{figure}[!htb]
\begin{minipage}[c]{0.3\columnwidth}
\includegraphics*[width= 1.8in]{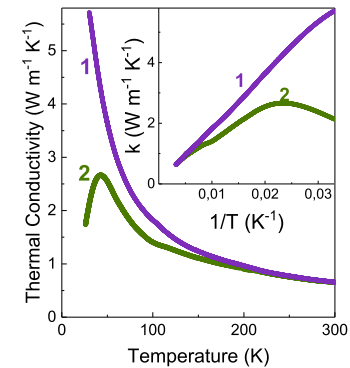}
\end{minipage}\hfill
  \begin{minipage}[c]{0.43\columnwidth}
   \caption{Temperature dependence of thermal conductivity $k$ for (Pb$_{0.98}$Ge$_{0.02}$)$_2$P$_2$S$_6$  (1) and \PPS (2) crystals. Inset: their $k(T^{-1})$ dependence. Data for $D$ is taken from [\onlinecite{b2019_50}].\label{fig8}}
  \end{minipage}
\end{figure}

For (Pb$_{0.98}$Ge$_{0.02}$)$_2$P$_2$S$_6$ crystal the thermal conductivity at low temperatures is bigger than in the case of pure \PPS crystal (see Fig.~\ref{fig8}). This is related to the induction of polar clusters of the ferroelectric phase when doping with Ge. The dielectric susceptibility of such clusters is smaller than the susceptibility of the paraelectric phase and the frequency of the lowest energy soft polar optic mode near the Brillouin zone (BZ) center is elevated. The growth of the soft optical mode frequency diminishes the probability of the optical phonon resonance scattering by acoustic phonons.\cite{b2019_32,b2019_33} At low temperatures heat is transferred by acoustic and lowest frequency optical phonons. Acoustic phonons with small wave numbers are involved mostly in normal scattering (N--process) that doesn’t contribute to thermal resistivity. The phonons from the optical branch near the BZ center also participate in Umklapp scattering (U--process) by lattice imperfections, which provide an effective thermal resistivity. So, the hardening of the optical branch lowers the population of the optical phonons and increases the thermal conductivity of (Pb$_{0.98}$Ge$_{0.02}$)$_2$P$_2$S$_6$ crystal at very low temperatures (see Fig.~\ref{fig8}). 
\begin{figure}[!htb]
\begin{minipage}[c]{0.3\columnwidth}
\includegraphics*[width= 1.8in]{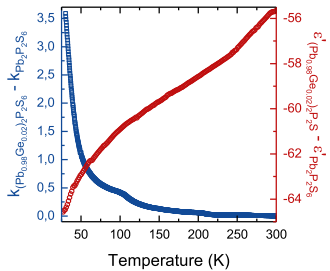}
\end{minipage}\hfill
  \begin{minipage}[c]{0.43\columnwidth}
   \caption{Temperature dependence of thermal conductivity $k$ and dielectric susceptibility $\varepsilon '$ difference for (Pb$_{0.98}$Ge$_{0.02}$)$_2$P$_2$S$_6$ and \PPS crystals. Data for $D$ is taken from [\onlinecite{b2019_50}].\label{fig9}}
  \end{minipage}
\end{figure}

Such explanation agrees with the comparison of the changes in the temperature dependencies of dielectric susceptibility and thermal conductivity induced by germanium (Fig.~\ref{fig9}). On cooling below 100~K, the difference in thermal conductivity between (Pb$_{0.98}$Ge$_{0.02}$)$_2$P$_2$S$_6$ and \PPS crystals rapidly rises, and oppositely --- dielectric susceptibility of (Pb$_{0.98}$Ge$_{0.02}$)$_2$P$_2$S$_6$ crystal quickly lowers relatively to \PPS crystal susceptibility. 

Such low temperature evolution of the dielectric susceptibility induced by germanium impurity reflects the hardening (frequency increase) of the lowest polar optic mode near the BZ center.  
\begin{figure}[!htb]
\includegraphics*[width=0.9\columnwidth]{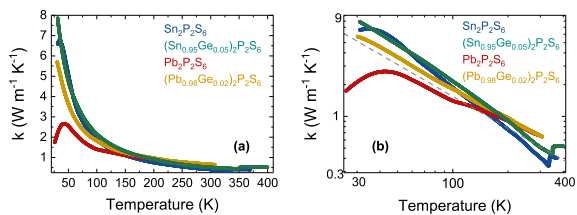}
 \caption{Temperature dependence of thermal conductivity $k$ for Sn$_2$P$_2$S$_6$, (Sn$_{0.95}$Ge$_{0.05}$)$_2$P$_2$S$_6$, \PPS and (Pb$_{0.98}$Ge$_{0.02}$)$_2$P$_2$S$_6$ in (a) normal coordinates and (b) in log-log scale. Grey dashed curve shows $k \sim T^{-1}$ behavior. Data for $D$ is taken from [\onlinecite{b2019_34,b2019_35,b2019_36,b2019_50}]. \label{fig10}}
\end{figure}

With the introduction of germanium into the lattice of \PPS crystal, the temperature dependence of the thermal conductivity coefficient $k(T)$ in a wide temperature range coincides with Eiken’s law, i.e. it is proportional to the inverse of temperature (see Fig.~\ref{fig10}). Such dependency gives evidence about the dominant role of three-phonon scattering processes in the thermal resistivity. In the case of \SPS ferroelectric phase, the introduction of Ge impurity also improves $k \sim T^{-1}$ temperature dependence for the thermal conductivity. 
\begin{figure}[!htb]
\includegraphics*[width=0.9\columnwidth]{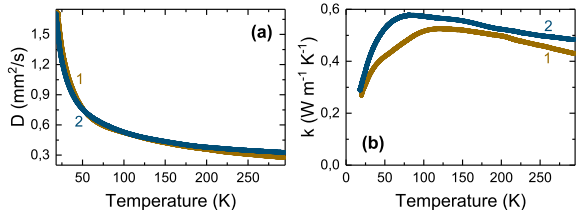}
 \caption{Temperature dependence of (a) thermal diffusivity $D$ and (b) thermal conductivity $k$ for (Pb$_{0.7}$Sn$_{0.25}$Ge$_{0.05}$)$_2$P$_2$S$_6$ (1) and (Pb$_{0.7}$Sn$_{0.25}$Ge$_{0.05}$)$_2$P$_2$Se$_6$ (2) mixed crystals. \label{fig11}}
\end{figure}

In the case of  (Pb$_{0.7}$Sn$_{0.25}$Ge$_{0.05}$)$_2$P$_2$S$_6$ mixed crystals the thermal conductivity temperature behavior (see Fig.~\ref{fig11}) is similar to observed in case of glassy materials, what demonstrates an effective phonon scattering in solid solutions with sublattice of mixed tin and lead cations. The addition of germanium impurity induces the dipole glass state, which appeared in the complex dielectric susceptibility frequency dependency below 100~K (see Fig.~\ref{fig4}). Only a small contribution to thermal conductivity by germanium addition is observed in the temperature range from 50 K to 120~K.  Similar behavior is also observed for (Pb$_{0.7}$Sn$_{0.25}$Ge$_{0.05}$)$_2$P$_2$Se$_6$ selenide solid solution.

\section{Discussion of results}
With the application of the local mode approach to the monoclinic ferroelectrics Sn$_2$P$_2$S$_6$, it was found\cite{b2019_4} a three-well potential energy surface. The nonlinear interaction of the vibration modes leads to this complex shape of local potential for spontaneous polarization fluctuations. Such nonlinearity is a result of significant electron-phonon interaction, that can be described as a second order Jahn-Teller effect related to the electron lone pair stereoactivity of Sn$^{2+}$ cations.\cite{b2019_4,b2019_15} The nonlinear lattice dynamics is reflected in the theoretically and experimentally observed\cite{b2019_6,b2019_20} complex nature of the soft mode related to continuous phase transition.

In the description of the microscopic origin of \SPS ferroelectric lattice instability, in addition to the second order Jahn-Teller effect, the $P^{4+} + P^{4+} \leftrightarrow P^{3+} + P^{5+}$ charge disproportionation was also considered.\cite{b2019_5} Such electronic correlations can be described within the presentation of Anderson’s electron pairs flipping, and thermodynamics of \SPS family ferroelectrics can be considered within the framework of BEG model.\cite{b2019_19,b2019_40,b2019_41} In this approximation, a change in the local three-well potential by flattening the side wells leads to a decrease of the calculated continuous phase transition temperature and a TCP is reached. Below TCP temperature, the first order ferroelectric phase transition line further drops down to 0~K. In the case of the family of \SPS ferroelectric crystals, such an evolution can be induced substituting tin by lead in mixed crystals \PSPS or under hydrostatic compression.\cite{b2019_5}

\begin{figure}[!htb]
\includegraphics*[width=0.9\columnwidth]{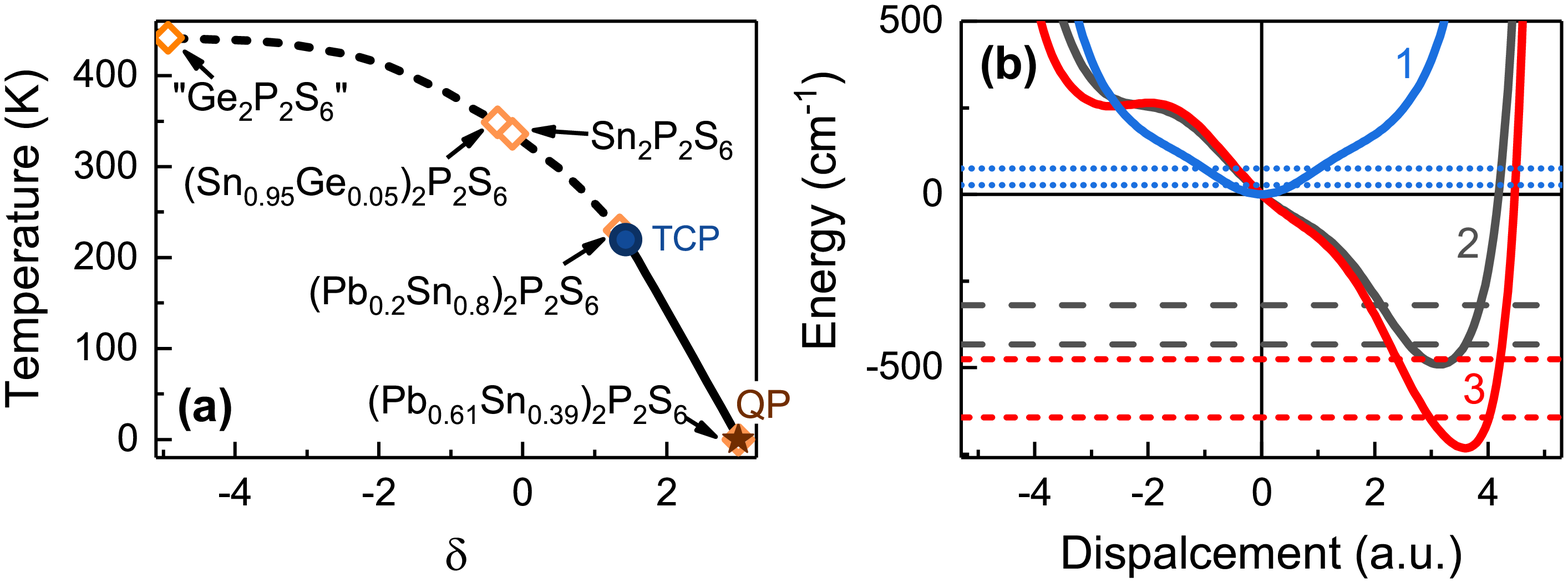}
 \caption{(a) the phase transition temperature as a function of $\delta = \Delta/J$, calculated in the mean-field approximation on the BEG model\cite{b2019_5} with shown points for “Ge$_2$P$_2$S$_6$”, (Sn$_{0.95}$Ge$_{0.05}$)$_2$P$_2$S$_6$, Sn$_2$P$_2$S$_6$, (Pb$_{0.2}$Sn$_{0.8}$)$_2$P$_2$S$_6$ and (Pb$_{0.61}$Sn$_{0.39}$)$_2$P$_2$S$_6$ crystals. Solid line denotes first order and dashed one second order phase transitions. Filled circle --- TCP, star --- QP. (b) the QAO local potentials at $T = 0$~K and their lowest energy levels for crystals Pb$_2$P$_2$S$_6$ (1), Sn$_2$P$_2$S$_6$ (2) and for virtual crystal “Ge$_2$P$_2$S$_6$” (3). \label{fig12}}
\end{figure}

The QAO model with the description of electronic recharging and lattice instability as pseudospin fluctuations in an anharmonic potential of three-well shape was proposed for a description of the temperature-pressure diagram of \SPS and of the temperature-composition diagram of \PSPS ferroelectric mixed crystals.\cite{b2019_5} In the QAO model, the real crystal lattice is represented as a system of one-dimensional interacting quantum anharmonic oscillators. A shape of the phase diagram calculated for the BEG model [see Fig.~\ref{fig12}(a)] correlates with the experimental observations. Here the on-site energy $\Delta$ changes with the variation of crystals chemical composition at almost constant intersite interaction $J$.  The dimensionless parameter $\delta = \Delta/J$  was estimated by using the following characteristics of the \PSPS mixed crystals $T - \delta$ phase diagram: the second order phase transition temperature for \SPS crystal, the coordinates of tricritical points on temperature-composition $T - y$ diagram, the composition $y$ at which the phase transition temperature goes down to zero.\cite{b2019_5,b2019_7} The shape of the local potential was determined with known values of on-site energy $\Delta$ [see Fig.~\ref{fig12}(b)]. 

The local potentials of quantum anharmonic oscillators for Pb$_2$P$_2$S$_6$, Sn$_2$P$_2$S$_6$ and virtual “Ge$_2$P$_2$S$_6$” crystals, which were determined by earlier\cite{b2019_5} described methodology, are shown at Fig.~\ref{fig12}(b). They are characterized by below listed values of zero-point energy $E_0 = \hbar\omega_0/2$, related frequency $\omega_0$ and temperature $T_x$:

Pb$_2$P$_2$S$_6$ -- $\omega_0\approx47$~cm$^{-1}$, $T_x\approx72$~K, $E_0 \approx0.003$~eV;

Sn$_2$P$_2$S$_6$ -- $\omega_0\approx60$~cm$^{-1}$, $T_x\approx86$~K, $E_0 \approx0.004$~eV;

Ge$_2$P$_2$S$_6$ -- $\omega_0\approx80$~cm$^{-1}$, $T_x\approx115$~K, $E_0 \approx0.005$~eV;

The shape of the local potential favors an off-center displacement of Sn$^{2+}$ cations in \SPS or Pb$^{2+}$ cations in \PPS crystal lattice that induces a local electric dipole. The local dipoles at a given inter-cell interaction $J$ cannot be ordered down to the lowest temperatures in case of \PPS crystal, but here the ferroelectric ground states may be reached via chemical substitution of lead cations by tin or germanium cations.

At low temperatures for ferroelectrics the quantum fluctuating electrical dipoles are coupled to the elastic steps of freedom. The quantum critical phase in three dimensional space $d$ is evident by the fact that the dielectric susceptibility depends on both the static and dynamic (frequency dependent) properties of the system, which results for multiaxial ferroelectrics, like perovskite SrTiO$_3$, in a unity rise of effective dimension --- $d_{eff} = d +1 = 4$.\cite{b2019_42,b2019_43} In the uniaxial ferroelectrics apart from short range interactions, the long range anisotropic electrical dipole interactions provide a further unity increase in the effective dimension to $d_{eff} = d +2 = 5$.\cite{b2019_44,b2019_45}

For SrTiO$_3$ below 25~K a nonmonotonic temperature dependence of dielectric susceptibility arises from optic and acoustic phonon coupling (electrostriction).\cite{b2019_25} The upturn in the inverse susceptibly occur when $T$ is less than 10\% of $T_x$, where $T_x$ is the temperature associated with the soft transverse optical phonon frequency $\omega$ at the BZ center in the zero-temperature limit. This means that fit of the dielectric susceptibility data to a quantum criticality model without taking into account of electrostrictive coupling is appropriate only for $T > 0.1T_x$. This condition is fulfilled for Pb$_2$P$_2$S$_6$ crystal where $0.1T_x \approx 7$~K.
\begin{figure}[!htb]
\includegraphics*[width=0.9\columnwidth]{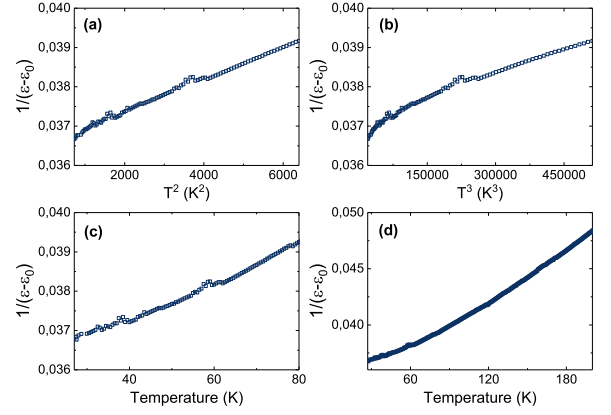}
 \caption{Reciprocal dielectric susceptibility of Pb$_2$P$_2$S$_6$ crystal as a function of temperature in different scales. \label{fig13}}
\end{figure}

We present the measured dielectric susceptibility $1/\varepsilon '(T)$ over the range 27--80~K for \PPS crystal in different temperature scales (see Fig.~\ref{fig13}). From the comparison of Fig.~\ref{fig13}(a) and Fig.~\ref{fig13}(b), it is seen that $1/\varepsilon '$ varies as $T^2$ in the region 27--80~K and doesn’t satisfy the $T^3$ quantum critical behavior. Above 80~K this crystal exhibits classical Curie-Weiss behavior [see Fig.~\ref{fig13}(c, d)]. We emphasize that for \PPS crystal the critical exponent is close to $\gamma = 2.0$, that is calculated and observed for multiaxial quantum critical systems like SrTiO$_3$,\cite{b2019_42,b2019_43}  and does not follow the theoretically predicted for uniaxial ferroelectrics value $\gamma = 3.0$\cite{b2019_44,b2019_45} which have been experimentally found in the case of BaFe$_{12}$O$_{19}$ and SrFe$_{12}$O$_{19}$ crystals.\cite{b2019_44}

The $1/T^2$ low temperature variation found for the dielectric susceptibility of \PPS close to the quantum critical point instead of the expected uniaxial behavior of $1/T^3$ can be explained at first glance by a screening phenomenon in semiconductor materials of the Sn(Pb)$_2$P$_2$S(Se)$_6$ system which weakens the electric dipole interaction. Such explanation is seen as appropriate for the above discussed critical behavior near the Lifshitz point in Sn$_2$P$_2$(Se$_{0.28}$S$_{0.72}$)$_6$ mixed crystal, which agrees with the theoretically predicted for systems with short-range interactions.\cite{b2019_35,b2019_36} But, for \PPS crystals at low temperatures, the electric conductivity is very small (below 10--14~Om$^{-1}$ cm$^{-1}$)\cite{b2019_46} and screening effects can’t be effective with a low concentration of the free charge carriers. 
\begin{figure}[!htb]
\includegraphics*[width=0.9\columnwidth]{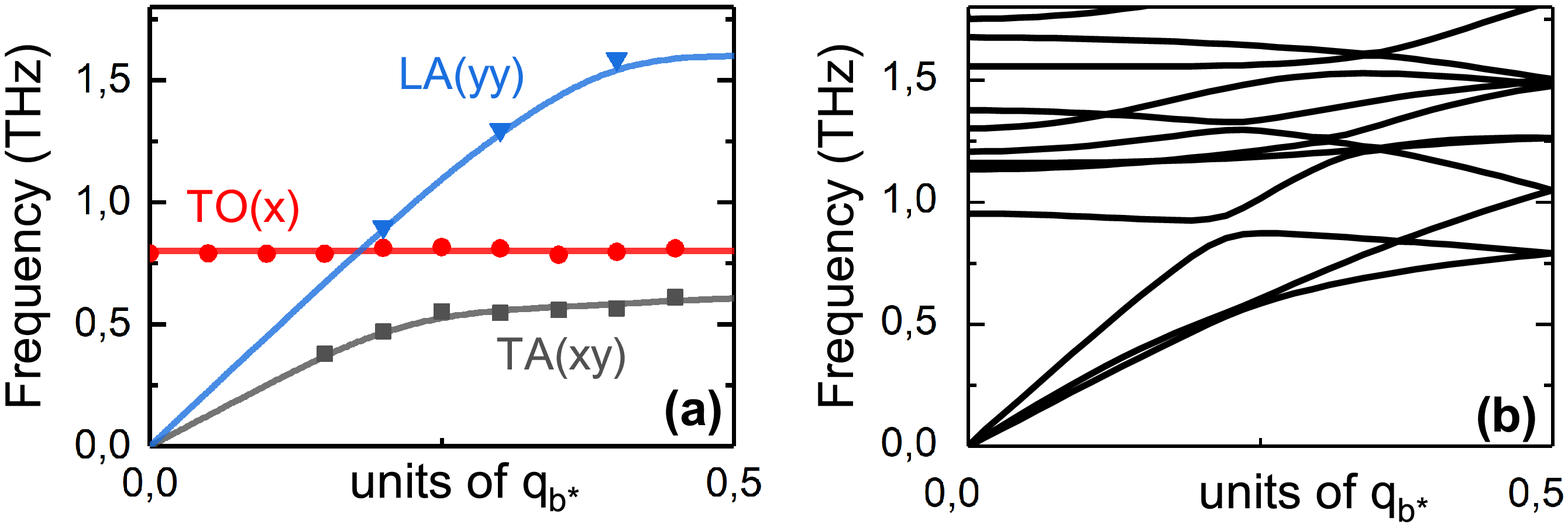}
 \caption{Acoustic and soft optic phonon branches: (a) determined by neutron scattering at 440~K in paraelectric P2$_1$/c phase of Sn$_2$P$_2$S$_6$;\cite{b2019_47} (b) calculated GGA approach of DFT for Pb$_2$P$_2$S$_6$.\cite{b2019_5} \label{fig14}}
\end{figure}

Figure~\ref{fig14} demonstrates, that for both \SPS and \PPS crystals the soft optic branch in the paraelectric phase is flat: soft phonons frequency slightly changes with the increase of wave number and moves from the BZ center to the edge, where the crossing with acoustic phonon branches occurs.\cite{b2019_47} On cooling to the continuous phase transition temperature $T_0 \approx 337$~K for \SPS crystal, in addition to the development of polar fluctuations near the BZ center, the antipolar fluctuations also strongly develop in the paraelectric phase. Here, critical behavior can be described as a crossover between Ising and XY universality classes, what is expected near bicritical points with coupled polar and antipolar order parameters and competing instabilities in the reciprocal wave vectors space.\cite{b2019_11} A similar situation obviously exists in the quantum paraelectric phase of \PPS crystal, where on cooling down to 0~K the flat optic phonon branch softens across wide reciprocal space in BZ. Besides, the long wavelength polar fluctuations grow together with a development of short wavelength antipolar fluctuations and, therefore, their nonlinear coupling can modify the quantum critical behavior.

\section{Conclusions}

The dipole ordering temperature of Sn(Pb)$_2$P$_2$S(Se)$_6$ materials may be tuned by chemical substitution realizing a ferroelectric quantum phase transition and quantum glassy or relaxor type phenomena in different parts of the phase diagram. The introduction of Ge impurity initiates several important phenomena: it increases the temperature of the phase transitions and improves the spontaneous polarization in the crystal; it doesn’t shift the coordinate of the Lifshitz point $x_{\textrm {LP}}$ in \SPSS mixed crystals; it initiates a more pronounced critical anomaly in \SPS crystals; it induces the appearance of a phase transition in the quantum paraelectric \PPS and inhomogeneous polar ordering in (Pb$_{0.7}$Sn$_{0.3}$)$_2$P$_2$S(Se)$_6$ crystals. The quantum fluctuations are destroyed in the mixed crystals, what follows from the comparison of the low temperature behavior of the thermal diffusivity and the complex dielectric susceptibility at different frequencies. 

By means of dielectric measurements it was shown that for \PPS crystal the real part of the dielectric susceptibility increases monotonously with decreasing temperature in the range from 300 K till 20~K. It was found that in the quantum critical regime the usual Curie-Weiss law of the inverse of dielectric susceptibility $1/\varepsilon '(T) \sim T$ changes into $1/\varepsilon '(T) \sim T^2$, which is the prominent criterion for quantum critical behavior. The nature of long-range dipole interactions in uniaxial materials predicts a dielectric susceptibility varying as $1/T^3$ close to the quantum critical point. But we found that the dielectric susceptibility varies as $1/ T^2$ as expected and observed in better known multi-axial systems. This result can be partially explained by a screening phenomenon in semiconductor materials of the Sn(Pb)$_2$P$_2$S(Se)$_6$ system which is effective at relatively high temperatures and weakens the electric dipole interactions. But due to the free charge carrier’s low concentration at low temperatures, evidently the nonlinear coupling between polar and antipolar fluctuation is surely related to the modification of the observed quantum critical behavior in \PPS crystal. 

The temperature dependence of the dielectric susceptibility has been analyzed in terms of the Barrett model that also demonstrates the presence of a quantum paraelectric state in \PPS type crystals. Small amounts of germanium impurity in (Pb$_{0.98}$Ge$_{0.02}$)$_2$P$_2$S$_6$ crystal induce the appearance of the ferroelectric phase, what is manifested in the decrease of the real part of the dielectric susceptibility below 75~K, deviating from the Barrett`s fit, and the appearance of a broad peak around 40~K. The observed behavior of the dielectric susceptibility temperature dependence demonstrates that (Pb$_{0.7}$Sn$_{0.3}$)$_2$P$_2$S$_6$ + 5\% Ge crystal doesn’t undergo a ferroelectric phase transition with polar ordering at macroscopic scale at any finite temperature, implying that a relaxor or dipole glass state appears below 50~K. The temperature dependence of the dielectric susceptibility at different frequencies for (Pb$_{0.7}$Sn$_{0.3}$)$_2$P$_2$Se$_6$ + 5\% Ge sample is similar to the observed one for the sulfide mixed crystal.

The thermal properties of Pb$_2$P$_2$S$_6$, (Pb$_{0.98}$Ge$_{0.02}$)$_2$P$_2$S$_6$, (Pb$_{0.7}$Sn$_{0.25}$Ge$_{0.05}$)$_2$P$_2$S$_6$  and (Pb$_{0.7}$Sn$_{0.25}$Ge$_{0.05}$)$_2$P$_2$Se$_6$ single crystals have been studied, as it was earlier performed for \SPS crystal doped by germanium,\cite{b2019_16} where Ge impurity sharpens the Ising type critical anomaly at the continuous ferroelectric transition. It was found that for (Pb$_{0.98}$Ge$_{0.02}$)$_2$P$_2$S$_6$ crystal the thermal conductivity at low temperature (near 50~K) is bigger than in the case of pure \PPS crystal. This is obviously related to the Ge induction of polar clusters of the ferroelectric phase. The dielectric susceptibility of such clusters is smaller than the susceptibility of the paraelectric phase and the frequency of the lowest energy soft optical mode near the BZ center is obviously elevated. The increase in the soft optical mode frequency diminishes the probability of acoustic phonons resonance scattering by optic phonons. At low temperatures, heat is transferred by the acoustic and the lowest frequency optical phonons. Acoustic phonons with the small wave numbers are involved mostly in normal scattering processes that don’t contribute to the thermal resistivity. The phonons from the optical branch near the BZ center also participate in Umklapp scattering by lattice imperfections which provides an effective thermal resistivity. Thus, the hardening of the optical branch lowers the population of the optical phonons and increases the thermal conductivity of (Pb$_{0.98}$Ge$_{0.02}$)$_2$P$_2$S$_6$ crystal. In the case of (Pb$_{0.7}$Sn$_{0.25}$Ge$_{0.05}$)$_2$P$_2$S(Se)$_6$ mixed crystals, the thermal conductivity behaves on cooling, like in glassy materials, which demonstrates an effective phonon scattering in solid solutions with sublattice of mixed tin and lead cations. Here germanium impurity induces the dipole glass state, which is manifested in the complex dielectric susceptibility frequency dependence below 100~K.


\begin{thebibliography}{10}
\bibitem{b2019_1}	M. M. Maior, M. I. Gurzan, Sh. B Molnar, I.P. Prits, and Yu. M. Vysochanskii, \href{https://doi.org/10.1109/58.852069}{IEEE Trans. of Ultrasonics. Ferroelectrics and Frequency Control \textbf{47}, 877 (2000)}.
\bibitem{b2019_2}	V. I. Litvinov, Izv. Akad. Nauk SSSR, Ser. Fiz. \textbf{51}, 1677 (1987).
\bibitem{b2019_3}J. Grigas, E. Talik, K. Glukhov, K. Fedyo, I. Stoika, M. Gurzan, I. Prits, A. Grabar, and Yu. Vysochanskii, \href{https://doi.org/10.1080/00150193.2011.578984}{ Ferroelectrics \textbf{418}, 134 (2001)}.
\bibitem{b2019_4}K. Z. Rushchanskii, Yu. M. Vysochanskii, and D. Strauch, \href{http://dx.doi.org/10.1103/PhysRevLett.99.207601}{Phys. Rev. Lett. \textbf{99}, 207601-1 (2007)}.
\bibitem{b2019_5}	R. Yevych, V. Haborets, M. Medulych, A. Molnar, A. Kohutych, A. Dziaugys, Ju. Banys, and Yu. Vysochanskii, \href{http://dx.doi.org/10.1063/1.4973005}{Low Temp. Phys. \textbf{42}, 1155 (2016)}.
\bibitem{b2019_6}	Yu. M. Vysochanskii, T. Janssen, R. Currat, R. Folk, J. Banys, J. Grigas, V. Samulionis, {\it Phase transitions in ferroelectric phosphorous chalcogenide crystals} (Vilnius University Publishing House, 2006).
\bibitem{b2019_7}	K. Z. Rushchanskii, R. M. Bilanych, A. A. Molnar, R. M. Yevych, A. A. Kohutych, S. I. Perechinskii, V. Samulionis, J. Banys, and Y. M. Vysochanskii, \href{http://dx.doi.org/10.1002/pssb.201552138}{Phys. Status Solidi B \textbf{253}, 384 (2016)}.
\bibitem{b2019_8} M. Blume, V. J. Emery, and R. B. Griffiths, \href{http://dx.doi.org/10.1103/PhysRevA.4.1071}{Phys. Rev. A. \textbf{4}, 1071 (1971)}.
\bibitem{b2019_9}W. Hoston and A. N. Berker, \href{http://dx.doi.org/10.1103/PhysRevLett.67.1027}{Phys. Rev. Lett. \textbf{67}, 1027 (1991)}.
\bibitem{b2019_10} Yu. M. Vysochanskii and V. Yu. Slivka, \href{http://dx.doi.org/10.1070/PU1992v035n02ABEH002217}{Sov. Phys. Usp. \textbf{35}, 123 (1992)}.
\bibitem{b2019_11}V. Liubachko, A. Oleaga, A. Salazar, R. Yevych, A. Kohutych, and Yu. Vysochanskii, \href{https://arxiv.org/abs/1912.13398}{arXiv:1912.13398 [cond-mat.mtrl-sci]} (2019).
\bibitem{b2019_12}K. Moriya, K. Iwauchi, M. Ushida, A. Nakagawa, K.Watanabe, S. Yano, S. Motojima, and Y. Akagi, \href{https://doi.org/10.1143/JPSJ.64.1775}{J. Phys. Soc. Jpn. \textbf{64}, 1775 (1995)}.
\bibitem{b2019_13}P. Ondrejkovic, M. Kempa, Y. Vysochanskii, P. Saint-Gregoire, P. Bourges, K. Rushchanskii, and J. Hlinka, \href{http://dx.doi.org/10.1103/PhysRevB.86.224106}{Phys. Rev. B \textbf{86}, 224106-1 (2012)}.
\bibitem{b2019_14}P. Ondrejkovic, M. Guennou, M. Kempa, Y. Vysochanskii, G. Garbarino, and J. Hlinka, \href{http://dx.doi.org/10.1088/0953-8984/25/11/115901}{J. Phys.: Condens. Matter \textbf{25}, 115901 (2013)}.
\bibitem{b2019_15} K. Glukhov, K. Fedyo, J. Banys, and Yu. Vysochanskii, \href{http://dx.doi.org/10.3390/ijms131114356}{Int. J. Mol. Sci. \textbf{13}, 14356 (2012)}.
\bibitem{b2019_16}V. Shvalya, A. Oleaga, A. Salazar, I. Stoika, and Yu. M. Vysochanskii, \href{http://dx.doi.org/10.1007/s10853-016-0091-5}{J. Mater. Sci. \textbf{51}, 8156 (2016)}.
\bibitem{b2019_36}A. Oleaga, V. Shvalya, A. Salazar, I. Stoika, and Yu. M. Vysochanskii, \href{http://dx.doi.org/10.1016/j.jallcom.2016.10.071}{J. Alloys Compd. \textbf{694}, 808 (2017)}.
\bibitem{b2019_17}U. V. Waghmare, N. A. Spaldin, H. C. Kandpal, and Ram Seshadri, \href{https://doi.org/10.1103/PhysRevB.67.125111}{Phys. Rev. B \textbf{67}, 125111 (2003)}; A. Walsh and G. W. Watson, \href{https://doi.org/10.1021/jp051822r}{J. Phys. Chem. B \textbf{109}, 18868 (2005)}
\bibitem{b2019_18} Makoto Naka, Hitoshi Seo, and Yukitoshi Motome, \href{https://doi.org/10.1103/PhysRevLett.116.056402}{Phys. Rev. Lett. \textbf{116}, 056402 (2016)}.
\bibitem{b2019_19}T. M. Rice and L. Sheddon, \href{http://dx.doi.org/10.1103/PhysRevLett.47.689}{Phys. Rev. Lett. \textbf{47}, 689 (1981)}.
\bibitem{b2019_20}R. Yevych, M. Medulych, and Yu. Vysochanskii, \href{https://doi.org/10.5488/CMP.21.23001}{Condens. Matter Phys. \textbf{21}, 23001-1 (2018)}.
\bibitem{b2019_21}I. Zamaraite, R. Yevych, A. Dziaugys, A. Molnar, J. Banys, S. Svirskas, and Yu. Vysochanskii, \href{http://dx.doi.org/10.1103/PhysRevApplied.10.034017}{Phys. Rev. Appl. \textbf{10}, 034017-1 (2018)}.
\bibitem{b2019_22}I. Zamaraite, A. Dziaugys, Yu. Vysochanskii, and J. Banys, Lithuanian Journal of Physics, 2020 (in press).
\bibitem{b2019_23}Yu. Vysochanskii, K. Glukhov, M. Maior, K. Fedyo, A. Kohutych, V. Betsa, I. Prits, and M. Gurzan, \href{https://doi.org/10.1080/00150193.2011.578979}{Ferroelectrics \textbf{418}, 124 (2011)}.
\bibitem{b2019_26}C. L. Wang and M. L. Zhao, \href{https://doi.org/10.1142/S2010135X1100029X}{J. Adv. Dielectr. \textbf{1}, 163 (2011)}.
\bibitem{b2019_25}S. E. Rowley, L. J. Spalek, R. P. Smith, M. P. M. Dean, M. Itoh, J. F. Scott, G. G. Lonzarich, and S. S. Saxena, \href{https://doi.org/10.1038/nphys2924}{ Nature Phys. \textbf{10}, 367 (2014)}.
\bibitem{b2019_42}Hideshi Fujishita, Shou Kitazawa, Masahiro Saito, Ryosuke Ishisaka, Hiroyuki Okamoto, and Toshihisa Yamaguchi, \href{https://doi.org/10.7566/JPSJ.85.074703}{J. Phys. Soc. Jpn. \textbf{85}, 074703 (2016)}.
\bibitem{b2019_27}N. Barman, P. Singh, C. Narayana, and K. B. R. Varma, \href{https://doi.org/10.1063/1.4973645}{AIP Adv. \textbf{7}, 035105 (2017)}.
\bibitem{b2019_28}J. G. Bednorz and K. A. Müller, \href{https://doi.org/10.1103/PhysRevLett.52.2289}{Phys. Rev. Lett. \textbf{52}, 2289 (1984)}.
\bibitem{b2019_29}Chen Ang and Zhi Yu, \href{https://doi.org/10.1103/PhysRevB.61.11363}{Phys. Rev. B \textbf{61}, 11363 (2000)}.
\bibitem{b2019_30}Yu. Vysochanskii, A. Molnar, R. Yevych, K. Glukhov, and M. Medulych, \href{https://doi.org/10.1080/00150193.2012.743850}{Ferroelectrics \textbf{440}, 31 (2012)}.
\bibitem{b2019_31}A. Oleaga, V. Shvalya, A. Salazar, I. Stoika, and Yu. M. Vysochanskii, \href{http://dx.doi.org/10.1016/j.jallcom.2016.10.071}{J. Alloys Compd. \textbf{694}, 808 (2017)}.
\bibitem{b2019_34}A. Oleaga, A. Salazar, M. Massot, and Yu. M. Vysochanskii, \href{http://dx.doi.org/10.1016/j.tca.2007.04.018}{Termochim. Acta \textbf{459}, 73 (2007)}.
\bibitem{b2019_35}A. Oleaga, A. Salazar, A. A. Kohutych, Yu. M. Vysochanskii, \href{http://dx.doi.org/10.1088/0953-8984/23/2/025902}{J. Phys.: Condens. Matter \textbf{23}, 025902-1 (2011)}.
\bibitem{b2019_49}K. Moriya, H. Kuniyoshi, K. Tashita, Y. Ozaki, S. Yano, and T. Matsuo, \href{http://dx.doi.org/10.1143/JPSJ.67.3505}{J. Phys. Soc. Jpn. \textbf{67}, 3505 (1998)}.
\bibitem{b2019_24}K. Moriya, T. Yamada, K. Saka, S. Yano, S. Baluya, T. Matsuo, P. Pritz, and Yu. Vysochanskii, \href{http://dx.doi.org/10.1023/A:1021643617039}{J. Therm. Anal. Calorim. \textbf{70}, 321 (2002)}.
\bibitem{b2019_48}V. Liubachko, A. Oleaga, A. Salazar, A. Kohutych, K. Glukhov, A. Pogodin, and Yu. Vysochanskii, \href{https://doi.org/10.1103/PhysRevMaterials.3.104415}{Phys. Rev. Materials \textbf{3}, 104415 (2019)}.
\bibitem{b2019_37}R. Folk, \href{http://dx.doi.org/10.1080/01411599908224501}{Phase Transitions \textbf{67}, 645 (1999)}.
\bibitem{b2019_32}Valentina Martelli, Julio Larrea Jiménez, Mucio Continentino, Elisa Baggio-Saitovitch, and Kamran Behnia, \href{https://doi.org/10.1103/PhysRevLett.120.125901}{Phys. Rev. Lett. \textbf{120}, 125901 (2018)}.
\bibitem{b2019_33}W. H. Huber, L. M. Hernandez, and A. M. Goldman, \href{https://doi.org/10.1103/PhysRevB.62.8588}{Phys. Rev. B \textbf{62}, 8588 (2000)}.
\bibitem{b2019_50}V. Shvalya, A. Oleaga, A. Salazar, A. Kohutych, and Yu. M. Vysochanskii,\href{https://doi.org/10.1166/mex.2017.1385}{	Mater. Express \textbf{7}, 361 (2017)}.
\bibitem{b2019_40}P. W. Anderson, \href{https://doi.org/10.1103/PhysRevLett.34.953}{Phys. Rev. Lett. \textbf{34}, 953 (1975)}.
\bibitem{b2019_41}A. Taraphder and P. Coleman, \href{https://doi.org/10.1103/PhysRevLett.66.2814}{Phys. Rev. Lett. \textbf{66}, 2814 (1991)}.
\bibitem{b2019_43}D. Rytz, U. T. Höchli, and H. Bilz, \href{https://doi.org/10.1103/PhysRevB.22.359}{Phys. Rev. B \textbf{22}, 359 (1980)}.
\bibitem{b2019_44}S. E. Rowley, Yi-Sheng Chai, Shi-Peng Shen, Young Sun, A. T. Jones, B. E. Watts, and J. F. Scott, \href{https://doi.org/10.1038/srep25724}{Sci. Rep. \textbf{6}, 25724 (2016)}.
\bibitem{b2019_45} S. E. Rowley, M. Hadjimichael, M. N. Ali, Y. C. Durmaz, J. C. Lashley, R. J. Cava, and J. F. Scott, \href{https://doi.org/10.1088/0953-8984/27/39/395901}{J. Phys.: Condens. Matter \textbf{27}, 395901 (2015)}.
\bibitem{b2019_46}V. M. Rizak, A. A. Bokotey, I. M. Rizak, K. Al'-Shoufi, and V. Yu. Slivka, \href{https://doi.org/10.1080/00150199708216183}{Ferroelectrics \textbf{192}, 149 (1997)}.
\bibitem{b2019_47}S. W. H. Eijt, R. Currat, J. E. Lorenzo, P. Saint-Grégoire, B. Hennion, and Yu. M. Vysochanskii, \href{https://doi.org/10.1007/s100510050431}{Eur. Phys. J. B \textbf{5}, 169 (1998)}.

\end{thebibliography}
\end{document}